\documentclass[10pt, conference]{IEEEtran}
\IEEEoverridecommandlockouts
\usepackage{cite}
\usepackage{amsmath,amssymb,amsfonts}
\usepackage{graphicx}
\usepackage{textcomp}
\usepackage{xcolor}
\usepackage{multirow}
\usepackage{subcaption}
\usepackage{tikz}
\usetikzlibrary{fit,positioning}
\usepackage{algorithm}
\usepackage{algpseudocode}
\usepackage{url}
\usepackage{hyperref}
\def\BibTeX{{\rm B\kern-.05em{\sc i\kern-.025em b}\kern-.08em
    T\kern-.1667em\lower.7ex\hbox{E}\kern-.125emX}}
\begin{document}

\title{Performance Analysis of QAOA Across Distributed Quantum Network Topologies Using SwitchQNet\thanks{This work was supported in part by JST [Moonshot R\&D Program] Grant Number [JPMJMS256K] and Grant Number [JPMJMS226C].}\\
}

\author{
\IEEEauthorblockN{
\begin{tabular}{ccc}
Samanvay Sharma &
Siyuan Niu &
Amin Taherkhani \\
\textit{Keio University} &
\textit{University of Central Florida} &
\textit{Keio University} \\
Fujisawa, Kanagawa, Japan &
Orlando, FL, USA &
Fujisawa, Kanagawa, Japan \\
samqc@keio.jp &
siyuan.niu@ucf.edu &
amin@sfc.wide.ad.jp
\end{tabular}
}

\vspace{0.8em}

\IEEEauthorblockN{
\begin{tabular}{cc}
Michal Hajdu\v{s}ek &
Rodney Van Meter \\
\textit{Keio University} &
\textit{Keio University} \\
Fujisawa, Kanagawa, Japan &
Fujisawa, Kanagawa, Japan \\
michal@sfc.wide.ad.jp &
rdv@sfc.wide.ad.jp
\end{tabular}
}
}

\maketitle

\begin{abstract}
Quantum data-center (QDC) architectures aim to scale distributed quantum computing (DQC) by interconnecting multiple quantum processing units (QPUs), but their performance depends strongly on how algorithmic communication patterns interact with entanglement generation, switch reconfiguration, and network topology. This paper studies the Quantum Approximate Optimization Algorithm (QAOA) as a graph-structured optimization workload for QDC-based distributed quantum computing. 
We adapt QAOA to SwitchQNet, a distributed quantum compiler framework that schedules communication and entanglement generation over switch-based QDC networks, by adding a routing generator that converts graph-dependent two-qubit cost interactions into remote-CX communication requests across QPUs. Using this extension, we evaluate QAOA instances across Clos, fat-tree, and spine-leaf topologies, measuring communication latency, EPR-pair overhead, EPR wait time, retry overhead, and sensitivity to buffer size, look-ahead depth, communication-qubit count, EPR latency, and EPR fidelity assumptions.
The results show that QAOA obtains modest but consistent latency reductions, highlighting its value as a diagnostic benchmark for studying the interaction between algorithm structure, entanglement management, and quantum-network architecture.
\end{abstract}

\begin{IEEEkeywords}
Quantum Networking, Distributed Quantum Computing, Quantum Approximate Optimization Algorithm
\end{IEEEkeywords}

% \section{Introduction}

\section{Introduction}

Quantum computing promises computational speedups for problem classes that are intractable on classical hardware, particularly in optimization, simulation, and cryptography\cite{nielsen2011quantum, divincenzo1995quantum, divincenzo2000physical}. However, today’s quantum processors remain constrained by device scale, noise, and connectivity in the Noisy Intermediate-Scale Quantum (NISQ) era\cite{preskill2018nisq}.
Software-level scale-up strategies such as quantum multiprogramming improve hardware utilization by co-scheduling multiple circuits on the same device~\cite{niu2022parallel,niu2023multiprogramming,ohkura2022simultaneous}. These techniques are valuable for near-term utilization, but they do not remove the architectural challenge of building systems with sufficient high-quality qubits, connectivity, and control resources for large-scale or fault-tolerant quantum computation (FTQC).
The need for quantum error correction (QEC) further increases this pressure since each logical qubit requires many physical qubits~\cite{devitt2013qec,fowler2012surface}.
Additionally, even if high-quality qubits exist, building a monolithic machine with the required number of physical qubits, connectivity, and control may be impractical in the near term. For instance, large monolithic systems may also introduce a single point of failure when system-level disruptions occur.

Distributed quantum computing (DQC) interconnects multiple smaller quantum processing  units (QPUs) using entanglement and classical coordination, trading monolithic integration for modular growth and allowing computation to scale to larger effective problem sizes by pooling resources across networked modules~\cite{vanmeter2006distributed,vanmeter2006architecture,vanmeter2014quantum,cacciapuoti2020quantum}. This direction is strengthened by continued progress in networked quantum systems, including experimental demonstrations of distributed computation across photonic links between separated modules\cite{vanmeter2012networking,vanmeter2014quantum,vanmeter2022architecture}. Recent quantum data-center (QDC) proposals specialize this idea to local, data-center-like environments where QPUs are organized into racks and connected through reconfigurable optical switch networks~\cite{shapourian2025qdc}. These architectures inherit useful ideas from classical interconnection networks, including Clos, fat-tree, and reconfigurable optical fabrics~\cite{clos1953nonblocking,leiserson1985fattrees,alfares2008scalable,avin2025reconfigurable,parsons2020optical}, but they also introduce quantum-specific constraints: Bell pairs/EPR pairs are consumed after use, have finite fidelity, may decohere while buffering, and require careful scheduling across communication qubits, Bell-state measurement resources, and switch configurations.

In parallel, data-center networking research has shown that optical switching may enable scalable, reconfigurable, high-bandwidth fabrics in classical infrastructure, an idea increasingly discussed as an enabling substrate for quantum data-center architectures that require frequent reconfiguration and low-contention connectivity~\cite{alfares2008scalable, avin2025reconfigurable}. Workload-aware DQC studies treat communication structure as a first-order compiler and architecture concern. Ferrari \emph{et al.}~\cite{ferrari2021compiler} characterize compilation overhead in distributed architectures, AutoComm~\cite{wu2022autocomm} and QuComm~\cite{wu2023qucomm} optimize recurring inter-QPU communication patterns, and DQC-QR jointly considers qubit placement and entanglement scheduling under network-resource and decoherence constraints~\cite{sundaram2025dqcqr}. At the architecture level, Q-Fly~\cite{sakuma2025qfly} combines a low-diameter optical interconnect with qubit placement and communication scheduling, while recent QDC benchmarking compares Q-Fly, BCube, Clos, and fat-tree architectures across circuit workloads~\cite{pouryousef2026benchmarking}. 
These trends motivate a need for systematic evaluation tools that jointly consider quantum program workload structures, entanglement generation, network topology, scheduling, resource allocation and sharing, and physical-layer constraints before large-scale quantum data centers exist as deployable hardware.

SwitchQNet~\cite{zhang2025switchqnet} is a recent compiler framework for QDCs that co-optimizes program-level communication demand with network-level entanglement generation. It uses look-ahead EPR scheduling, collective in-rack EPR generation, cross-rack EPR splitting, distillation, and retry mechanisms to reduce communication latency while controlling fidelity and congestion overheads. The original SwitchQNet evaluation demonstrates substantial latency reductions for benchmarks including Multi-Control Target (MCT)~\cite{asano2005mct,barenco1995mct}, Quantum Fourier Transform (QFT)~\cite{coppersmith2002approximate}, Grover~\cite{grover1996fast} and Ripple Carry Adder (RCA)~\cite{cuccaro2004rca}. 

The Quantum Approximate Optimization Algorithm (QAOA) is a
hybrid variational algorithm for combinatorial optimization.
Given a classical objective function, QAOA encodes its value in a
cost Hamiltonian $H_C$. For weighted MaxCut on a graph
$G=(V,E)$, a common formulation is
\[
H_C=\sum_{(i,j)\in E}w_{ij}\frac{I-Z_iZ_j}{2},
\]
whose computational-basis eigenvalue equals the weight of the
corresponding cut. Starting from the uniform superposition
$|+\rangle^{\otimes n}$, QAOA alternates the cost unitary
$U_C(\gamma_\ell)=e^{-i\gamma_\ell H_C}$ with a mixer unitary
$U_M(\beta_\ell)=e^{-i\beta_\ell H_M}$ for $p$ layers. A
classical optimizer updates the resulting $2p$ parameters to
maximize the expected objective value, and measurements of the
final state produce candidate solutions~\cite{farhi2014qaoa}.

For MaxCut, each term of $H_C$ acts on the two qubits associated with one problem-graph edge and can be implemented as an edge-local $ZZ$ rotation, commonly decomposed into a CX--$R_z$--CX sequence. The standard mixer consists only of local single-qubit rotations. Consequently, after logical qubits are assigned to QPUs, the graph edges whose endpoints lie on different QPUs identify the cost interactions requiring inter-QPU communication. This provides a transparent relationship between the problem instance, qubit placement, and the initial communication demand. The QDC network topology and scheduler then determine whether these requests use in-rack or cross-rack entanglement, as well as the latency and entanglement-resource cost of executing them. Although these mapping and topology effects apply to all distributed circuits, QAOA provides a controlled benchmark whose interaction pattern can be varied through the problem graph while holding other properties, such as the qubit count and circuit depth, fixed.

This work adapts QAOA to SwitchQNet by adding a routing generator that converts graph-edge cost interactions into remote-CX communication requests. We evaluate the resulting workload using SwitchQNet's existing QDC scheduling and entanglement-management mechanisms across multiple network topologies and hardware and compiler parameters. The objective is to broaden SwitchQNet's benchmark coverage and characterize how QAOA's graph-defined communication pattern influences latency, EPR overhead, EPR wait time, retry overhead, and QEC-enabled execution.

\section{Solution Methodology}
\label{sec:methodology}

We extend the SwitchQNet workflow by adding QAOA as a benchmark workload. The extension converts QAOA’s graph-dependent two-qubit cost interactions into SwitchQNet-compatible remote-CX routing requests between QPUs, allowing QAOA to be evaluated under the same routing, buffering, and entanglement-management assumptions as the existing benchmarks.

\subsection{QAOA Routing Generation}

We model QAOA on a problem graph $G=(V,E)$, where each vertex corresponds to a logical qubit and each edge $(u,v)\in E$ induces a two-qubit cost interaction. For an Ising- or MaxCut-style QAOA cost unitary, each edge interaction can be decomposed into a CX--$R_z$--CX pattern. Since $R_z$ rotations and mixer operations are single-qubit gates, they are local to each QPU and do not generate inter-QPU communication requests. Therefore, only the two CX operations associated with an edge require routing when the corresponding qubits are mapped to different QPUs.

For each QAOA layer, the routing generator scans the edge list $E$. For an edge $(u,v)$, qubits are mapped to QPUs by integer division over the configured qubits per QPU. If $u$ and $v$ map to different QPUs, the generator appends two remote-CX requests, $[Q_u,Q_v,\mathrm{CX}]$, corresponding to the CX--$R_z$--CX decomposition of the edge cost interaction. The $R_z$ rotation and mixer operations are local and do not generate SwitchQNet routing requests. We evaluate one QAOA layer ($p=1$) at the routing level. The classical parameter-optimization loop, quantum-state evolution, and solution quality are outside the scope of this study.

\subsection{Graph Instance Generation}
To study how QAOA communication depends on the problem structure, we implement $k$-regular graph instances. For a $k$-regular instance, every logical qubit participates in exactly $k$ edge-cost terms, providing a uniform per-qubit interaction degree. After qubit placement, only edges whose endpoints reside on different QPUs generate remote-CX requests.
Random and all-to-all graph generators are also added to the benchmark program to support future experiments with different communication densities.
For our experiments, we write regular QAOA graph instances generated by a pairing heuristic with target degree $k=3$. Candidate self-loops (nodes connecting to themselves) are discarded and duplicate edges (redundant node connections like (0,1) and (1, 0)) are merged before routing. 
The final edge list is passed directly to the QAOA routing generator.

\subsection{Integration with SwitchQNet Evaluation}

After generating the QAOA routing list, we execute it through the same SwitchQNet evaluation pipeline used for the original benchmark programs. The hardware configuration defines the QDC architecture, including the network topology (e.g., Clos, fat-tree, or spine-leaf), the number of racks, the number of QPUs per rack, the number of qubits per QPU, and communication-specific resources such as dedicated communication qubits and EPR buffer capacity. Given this configuration, the scheduler processes the routing list as a sequence of remote-CX requests and maps each request onto available network resources. It first determines whether the interaction occurs within a rack or across racks, distinguishing lower-latency, high-fidelity in-rack entanglement from slower and lower-fidelity cross-rack entanglement. The scheduler then allocates communication qubits, schedules EPR generation and consumption, manages buffer usage, and applies routing decisions based on the selected scheduling strategy, producing a time-ordered execution that reflects both network constraints and entanglement availability.

This integration does not modify SwitchQNet's underlying scheduling mechanisms. Instead, QAOA is added as an additional workload that exercises the existing scheduling and entanglement-management strategies. These include look-ahead scheduling, which reserves entanglement resources in advance based on future communication demand; collective in-rack EPR generation, where high-fidelity entangled pairs are generated locally within racks; cross-rack EPR splitting, which extends entanglement across racks via intermediate nodes; distillation, which improves the fidelity of noisy EPR pairs; and retry logic, which reattempts failed entanglement generation. The resulting outputs are processed using the same metric pipeline as the original SwitchQNet experiments~\cite{zhang2025switchqnet}, including communication latency, EPR-pair counts, EPR overhead, wait time, and retry overhead. This allows QAOA to be compared directly with the existing SwitchQNet benchmarks while isolating the effect of its graph-dependent communication pattern.

The implementation is available online~\cite{switchqnet_code}.

\section{Framework Setup}

\subsection{Architecture Setup}

\begin{figure}[t]
\centering
\scriptsize

\resizebox{0.5\textwidth}{!}{%
\begin{tikzpicture}[
    qpu/.style={circle, draw, minimum size=4mm, inner sep=0pt},
    tor/.style={rectangle, draw, rounded corners, minimum width=7mm, minimum height=4mm},
    sw/.style={rectangle, draw, minimum width=7mm, minimum height=4mm},
    upl/.style={rectangle, draw, minimum width=7mm, minimum height=4mm},
    link/.style={thin},
    uplink/.style={thin},
    rackbox/.style={draw, dashed, inner sep=3pt},
    racklabel/.style={font=\scriptsize}
]

\def\racksep{2.6}
\def\racky{1.6}
\def\aggy{3.0}
\def\corey{4.3}

\newcommand{\rack}[5]{
  \node[tor] (#1tor) at (#2,#3) {#5};
  \node[qpu] (#1q1) at (#2-0.38,#3-0.50) {};
  \node[qpu] (#1q2) at (#2+0.38,#3-0.50) {};
  \node[qpu] (#1q3) at (#2-0.38,#3-0.95) {};
  \node[qpu] (#1q4) at (#2+0.38,#3-0.95) {};
  \foreach \q in {q1,q2,q3,q4}{
    \draw[link] (#1tor) -- (#1\q);
  }
  \node[rackbox, fit=(#1tor)(#1q1)(#1q2)(#1q3)(#1q4)] (#1box) {};
  \node[racklabel, below=2pt of #1box] {Rack #4};
}

\node[upl] (c4) at (1.0*\racksep,\corey) {4};
\node[upl] (c5) at (2.0*\racksep,\corey) {5};

\node[sw] (a6) at (0.5*\racksep,\aggy) {6};
\node[sw] (a7) at (1.5*\racksep,\aggy) {7};
\node[sw] (a8) at (2.5*\racksep,\aggy) {8};
\node[sw] (a9) at (3.5*\racksep,\aggy) {9};

\rack{r0}{0*\racksep}{\racky}{0}{0}
\rack{r1}{1*\racksep}{\racky}{1}{1}
\rack{r2}{2*\racksep}{\racky}{2}{2}
\rack{r3}{3*\racksep}{\racky}{3}{3}

\foreach \c in {c4,c5}{
  \foreach \a in {a6,a7,a8,a9}{
    \draw[uplink] (\c) -- (\a);
  }
}

\draw[uplink] (r0tor) -- (a6);
\draw[uplink] (r0tor) -- (a7);
\draw[uplink] (r1tor) -- (a6);
\draw[uplink] (r1tor) -- (a7);

\draw[uplink] (r2tor) -- (a8);
\draw[uplink] (r2tor) -- (a9);
\draw[uplink] (r3tor) -- (a8);
\draw[uplink] (r3tor) -- (a9);

\node[tor] at (4.25*\racksep,\corey) {};
\node[anchor=west] at (4.38*\racksep,\corey) {ToR};
\node[qpu] at (4.25*\racksep,\corey-0.45) {};
\node[anchor=west] at (4.38*\racksep,\corey-0.45) {QPU};
\node[sw] at (4.25*\racksep,\corey-0.90) {};
\node[anchor=west] at (4.38*\racksep,\corey-0.90) {Switch};

\end{tikzpicture}
}

\caption{Clos topology for $n=4$ generated from \texttt{gen\_clos\_conn} function. Each rack contains one ToR switch and four QPUs.}
\label{fig:clos_topology}
\vspace{-2em}
\end{figure}
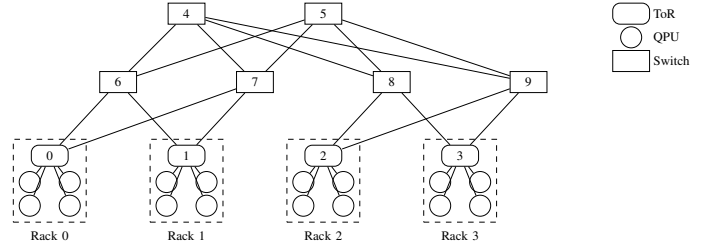

We evaluate SwitchQNet using a detailed event-driven simulation of a modular quantum computing architecture. Clos~\cite{pouryousef2026benchmarking, shapourian2025qdc, clos1953nonblocking} is used as the default topology\textit{(Fig. \ref{fig:clos_topology})}, with fat-tree and spine-leaf included for topology comparison, consistent with the architectural assumptions adopted in prior SwitchQNet experiments. 
As also shown in Figure \ref{fig:clos_topology}, a Clos network consists of an ingress switch stage at the bottom that receives incoming data (also called top-of-rack (ToR) switches), a middle stage that interconnects all switches, and an egress switch stage at the top that delivers data traffic to destination ports.
Fat-trees in SwitchQNet follow a standard implementation where modern data centers use three layers of switches: an edge/access layer at the bottom where servers are connected (ToR switches), an aggregation layer in the middle which interconnects groups of edge switches, and a core layer at the top providing connectivity between all aggregation switches.
The two-tier spine-leaf architecture for our case may be viewed as a "folded" three-stage Clos network where the ingress and egress layers merge into a single layer as "leaf-switches" to perform the function of both ingress and egress layers, and may also be visualized as a "two-layer" fat-tree.

Each rack is equipped with a rack-level router supporting four Bell-state measurement (BSM) units, enabling concurrent entanglement swapping operations. Cross-rack connections are modeled with higher routing weights than in-rack links to capture the increased latency and resource cost associated with long-distance entanglement generation. 

For consistency with prior work~\cite{zhang2025switchqnet}, we adopt the same experimental configuration as the SwitchQNet framework and focus on a representative subset of program instances for evaluation. 
The look-ahead depth (10), latencies of in-rack \& cross-rack EPR pair generation (~0.1 ms and 10 ms respectively), and switch reconfiguration latency (~1 ms) are set according to SwitchQNet's listed numbers when estimating rates and fidelity of in-rack and cross-rack EPR pair generation.
The program size is given by the total number of data qubits involved in the compilation program, for example, \textit{QAOA-480} represents a collection of 4 racks with 4 QPUs per rack and 30 data qubits per QPU, which total up to $4 \times 4 \times 30 = 480$ data qubits. For general reference, in each QPU, two qubits are reserved for communication, while the remaining qubits are allocated to computation and local caching. Communication qubits are shared between inter-QPU entanglement generation and cache reservation, reflecting hardware constraints. All of the used configurations are under Table~\ref{tab:qaoa_configs}.

\begin{table}[t]
\caption{QAOA Workload Configurations}
\label{tab:qaoa_configs}
\centering
\footnotesize
\begin{tabular}{lccc}
\hline
\textbf{Workload} & \textbf{Racks} & \textbf{QPUs/Rack} & \textbf{Qubits/QPU} \\
\hline
QAOA-480 & 4  & 4 & 30 \\
% QAOA-240 & 4  & 3 & 20 \\
QAOA-360 & 4  & 3 & 30 \\
QAOA-540 & 9  & 3 & 20 \\
QAOA-600 & 4  & 5 & 30 \\
QAOA-608 & 4  & 4 & 38 \\
QAOA-720 & 4  & 6 & 30 \\
QAOA-960 & 16 & 3 & 20 \\
Fat-tree QAOA-960 & 8 & 3 & 30 \\
Spine-leaf QAOA-720 & 6 & 4 & 30 \\
\hline
\end{tabular}
\vspace{-2em}

\end{table}

Different program numbers represent different numbers and combinations of data qubits, QPUs and racks. These different program configurations are selectively used to study scaling behavior. Unless otherwise stated, \textit{QAOA-480} is used as the baseline configuration for benchmarking and comparison across different workloads. When using fat-tree topology, we use QAOA-960 with 8 racks, 3 QPUs/rack, \& 30 qubits/QPU, and when using spine-leaf topology, we use QAOA-720 with 6 racks, 4 QPUs/rack, \& 30 qubits/QPU.

We use the same latency assumptions for cross-rack and in-rack EPR pair generation as well as switch reconfiguration that occur during inter-QPU communication, as mentioned in the source SwitchQNet paper~\cite{zhang2025switchqnet}. All time units are in milliseconds (ms) unless stated otherwise. Scheduling times wherever mentioned are in seconds (s). Experiments were run on macOS 15.5 using Python on a MacBook Pro with a 2.3 GHz 8-core Intel Core i9 CPU and 32 GB memory.

\subsection{Benchmark Programs}

The pre-existing set of benchmark programs, MCT, QFT, RCA, and Grover's algorithm, are retained for comparison. We add QAOA as our choice of benchmark quantum program to evaluate routing behavior under non-trivial communication patterns. Within the target-degree-3 QAOA workload, only the problem unitary requires two-qubit operations across logical qubits, while the mixer unitary consists solely of single-qubit gates and does not incur inter-QPU communication. As a result, the qubits are connected as a circuit in the framework laid out by the QAOA routing model, and integrated into the existing SwitchQNet benchmark framework using the same circuit abstraction and routing interface as the other programs. This design allows QAOA to be evaluated alongside existing benchmarks without modifying the underlying routing or scheduling infrastructure, ensuring methodological consistency across experiments.

\subsection{Metrics}

To evaluate the compilation performance of the introduced benchmarking programs in SwitchQNet, we use the following metrics:

The first is the overall communication latency of the benchmarking programs, termed as 'latency', normalized by switch reconfiguration latency. These are all considered under inter-QPU communication while intra-QPU computation time is neglected because it is much shorter than inter-QPU communication time under the adopted model.

The second metric is the fidelity overhead, specifically the EPR overhead which is the number of additional distilled in-rack EPR pairs required arising due to communication splits, represented in terms of percentages of the original in-rack EPR pair numbers and evaluated in a weighted manner.

There is also an additional fidelity metric which is the average wait time of EPR pairs waiting in the buffer normalized by switch reconfiguration latency as well, but since the impacts of wait time are more dependent upon the QPU hardware in use, specifically on the coherence of the qubits involved in computation, not something that would be considered in network compilation here, we consider these separately and consider the EPR overhead percentage as the main contributor to fidelity effects here.

The third metric will be the retry overhead which shows the overhead for compilation time added by retry counts:

\begin{equation}
    \text{Retry Overhead} = \frac{t_{\text{attempts}}}{t_{\text{compilation}}},
\end{equation}

where $t_{\text{attempts}}$ is the total attempted time step number in the compilation and $t_{\text{compilation}}$ is the actual number of time steps in the compilation results.

Based on these metrics, we shall be able to observe and assess the workload impacts of running QAOA on choice of network setups.

\subsection{Comparison Baseline}

We use the same baseline as integrated by SwitchQNet for all benchmark tests, which is based on the QuComm compiler framework for general DQC communication \cite{wu2023qucomm}, matching the preprocessing assumptions used in SwitchQNet. The EPR pair list is acquired by each of the benchmarking algorithms with pair numbers minimized through a buffer-aware compilation step equivalent to SwitchQNet, followed by the buffer-assisted on-demand approach of SwitchQNet.

\section{Experimental Results}

\begin{table*}[h]
\caption{QAOA performance across different architecture settings. Latency and wait time are normalized by the switch reconfiguration latency, and are therefore reported in dimensionless time units corresponding to multiples of a switch reconfiguration cycle.}
\label{tab:qaoa_results}
\begin{center}
\scriptsize
\setlength{\tabcolsep}{2.6pt}
\renewcommand{\arraystretch}{0.96}
\resizebox{0.98\textwidth}{!}{%
\begin{tabular}{|l|l|r|r|c|r|r|r|r|r|r|r|}
\hline
\textbf{Experiment} & \textbf{Benchmark} & \multicolumn{1}{c|}{\textbf{\shortstack{Baseline\\Latency}}} & \multicolumn{1}{c|}{\textbf{\shortstack{Ours\\Latency}}} & \textbf{\shortstack{Improv.\\Factor}} & \multicolumn{1}{c|}{\textbf{\shortstack{\#cross-rack\\EPR}}} & \multicolumn{1}{c|}{\textbf{\shortstack{\#in-rack\\EPR}}} & \multicolumn{1}{c|}{\textbf{\shortstack{\#dist.\\EPR}}} & \multicolumn{1}{c|}{\textbf{\shortstack{Baseline\\Wait}}} & \multicolumn{1}{c|}{\textbf{\shortstack{Ours\\Wait}}} & \multicolumn{1}{c|}{\textbf{\shortstack{Addit.\\Wait}}} & \multicolumn{1}{c|}{\textbf{\shortstack{Retry\\Overh.}}} \\
\hline

\multirow{4}{*}{Increase \#QPUs/rack}
& QAOA-360 & 18,438 & 9,835 & $\boldsymbol{1.87\times}$ & 782 & 192 & 409 & 0.68 & 19.60 & 18.92 & 1.00 \\
\cline{2-12}
& QAOA-480 & 20,704 & 10,363 & $\boldsymbol{2.00\times}$ & 1,064 & 282 & 737 & 0.56 & 17.82 & 17.25 & 1.00 \\
\cline{2-12}
& QAOA-600 & 22,992 & 10,584 & $\boldsymbol{2.17\times}$ & 1,348 & 352 & 1,028 & 0.47 & 16.38 & 15.91 & 1.00 \\
\cline{2-12}
& QAOA-720 & 21,650 & 11,051 & $\boldsymbol{1.96\times}$ & 1,604 & 456 & 1,237 & 0.45 & 15.93 & 15.48 & 1.04 \\
\hline

\multirow{3}{*}{Increase \#qubits/QPU}
& QAOA-480 & 20,704 & 10,363 & $\boldsymbol{2.00\times}$ & 1,064 & 282 & 737 & 0.56 & 17.82 & 17.25 & 1.00 \\
\cline{2-12}
& QAOA-608 & 26,952 & 12,515 & $\boldsymbol{2.15\times}$ & 1,356 & 352 & 1,048 & 0.63 & 17.86 & 17.23 & 1.00 \\
\cline{2-12}
& QAOA-720 & 30,582 & 14,503 & $\boldsymbol{2.11\times}$ & 1,604 & 440 & 1,397 & 0.54 & 19.88 & 19.34 & 1.00 \\
\hline

\multirow{3}{*}{Increase \#racks}
% & QAOA-240$^*$ & 13,543 & 8,082 & $\boldsymbol{1.68\times}$ & 530 & 122 & 191 & 0.49 & 12.52 & 12.03 & 1.04 \\
% \cline{2-12}
& QAOA-540 & 18,196 & 9,231 & $\boldsymbol{1.97\times}$ & 1,430 & 126 & 701 & 0.34 & 17.25 & 16.91 & 1.00 \\
\cline{2-12}
& QAOA-960 & 20,000 & 13,818 & $\boldsymbol{1.45\times}$ & 2,706 & 124 & 843 & 0.11 & 13.66 & 13.55 & 1.10 \\
\hline

Clos top. (unchanged) & QAOA-480 & 22,695 & 16,722 & $\boldsymbol{1.36\times}$ & 1,064 & 282 & 84 & 0.78 & 15.87 & 15.08 & 1.00 \\
\hline
Fat-tree top. & QAOA-960 & 30,604 & 21,782 & $\boldsymbol{1.41\times}$ & 2,532 & 266 & 203 & 0.30 & 12.61 & 12.31 & 1.00 \\
\hline
Spine-leaf top. & QAOA-720 & 27,249 & 19,142 & $\boldsymbol{1.42\times}$ & 1,788 & 272 & 112 & 0.45 & 13.05 & 12.60 & 1.00 \\
\hline

\end{tabular}}
\end{center}
\vspace{-2em}

\end{table*}
\subsection{Primary Experiment Results with QAOA}

The goal of these experiments is to evaluate how network-aware compilation and scheduling affect the performance of a structured optimization algorithm when key architectural parameters are varied. The results are summarized in Table~\ref{tab:qaoa_results}.

Across all experiments, the extended SwitchQNet workflow consistently outperforms the baseline compiler in terms of end-to-end latency for QAOA, although the magnitude of improvement is smaller than that observed for communication-intensive benchmarks such as Grover, QFT, and RCA. This behavior is expected given QAOA’s graph-dependent interaction pattern. When increasing the number of QPUs per rack, QAOA exhibits improvement factors ranging from approximately \(1.9\times\) to \(2.2\times\), indicating that SwitchQNet’s look-ahead scheduling and collective in-rack EPR generation are effective at hiding switch reconfiguration latency even for optimization workloads. Notably, the improvement factor does not increase monotonically with system size, reflecting the fact that QAOA’s communication demand grows with problem structure rather than purely with hardware scale.

The QAOA profile increases the relative share of cross-rack communication in the evaluated workload mix, reinforcing the original SwitchQNet observation that cross-rack EPR generation dominates latency despite representing a smaller fraction of total EPR operations.

When increasing the number of qubits per QPU, QAOA maintains stable improvement factors slightly above \(2\times\). This suggests that increasing local computational capacity without proportionally increasing inter-QPU communication does not significantly change the relative benefit of network-aware scheduling. In this regime, QAOA remains primarily limited by cross-rack entanglement availability rather than local gate execution, and SwitchQNet continues to reduce idle waiting time by overlapping EPR generation with computation.

As the system expands to larger rack counts, QAOA’s improvement factor remains modest, typically between \(1.4\times\) and \(2.0\times\). While absolute latency increases with additional racks due to higher cross-rack communication cost, SwitchQNet still achieves measurable gains over the baseline. The results indicate that QAOA benefits from network-aware scheduling even in more distributed settings, but that its limited communication density constrains the achievable latency reduction compared to benchmarks with heavier cross-rack interaction.

The topology comparison further reinforces this observation: QAOA is relatively insensitive to the specific topology choice compared to other benchmarks, provided that sufficient path diversity exists. As a result, QAOA serves as a useful contrast workload that exposes how algorithm structure influences the effectiveness of network-level optimizations.

Overall, these results demonstrate that while QAOA does not achieve the dramatic improvement factors observed for more communication-heavy programs, it consistently benefits from SwitchQNet’s scheduling strategies across all evaluated dimensions. This confirms QAOA’s suitability as both a representative application workload and a diagnostic benchmark for studying the interaction between distributed quantum algorithms and network architectures.

\begin{figure*}[h]
    \centering
        \includegraphics[width=0.9\textwidth]{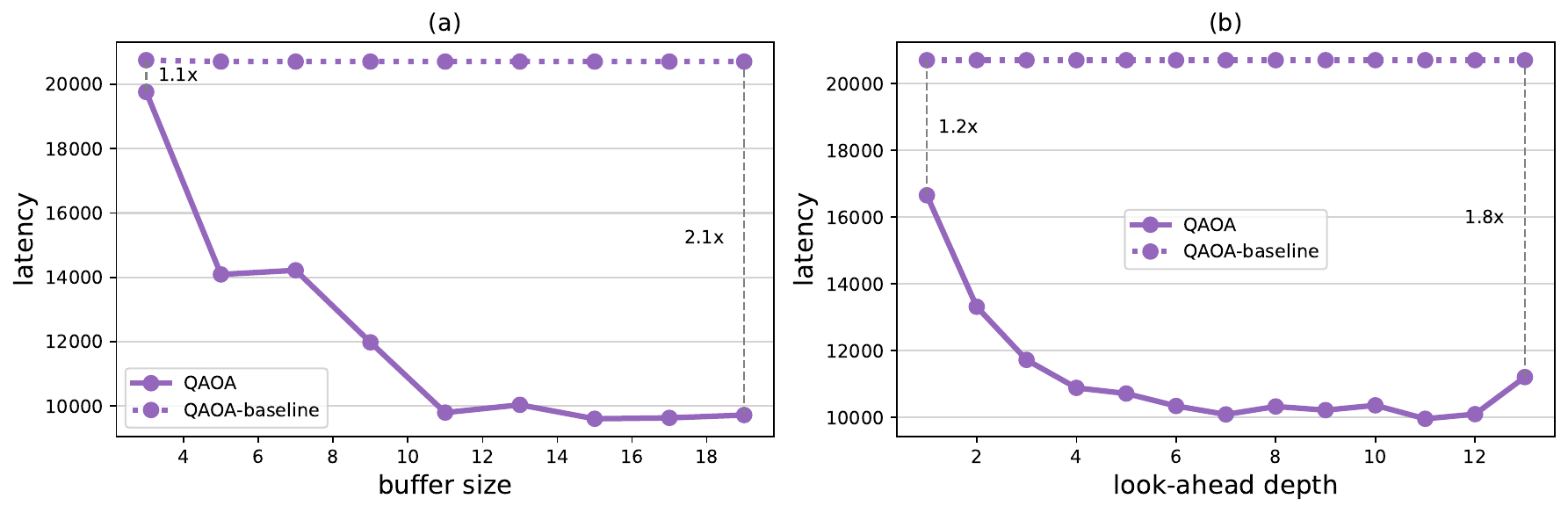}
        \caption{QAOA performance improvement of SwitchQNet varying with (a) buffer-size and (b) look-ahead depth}
        \label{fig:fig8_qaoa}
\end{figure*}

 \begin{figure*}[h]
        \centering
        \includegraphics[width=0.9\textwidth]{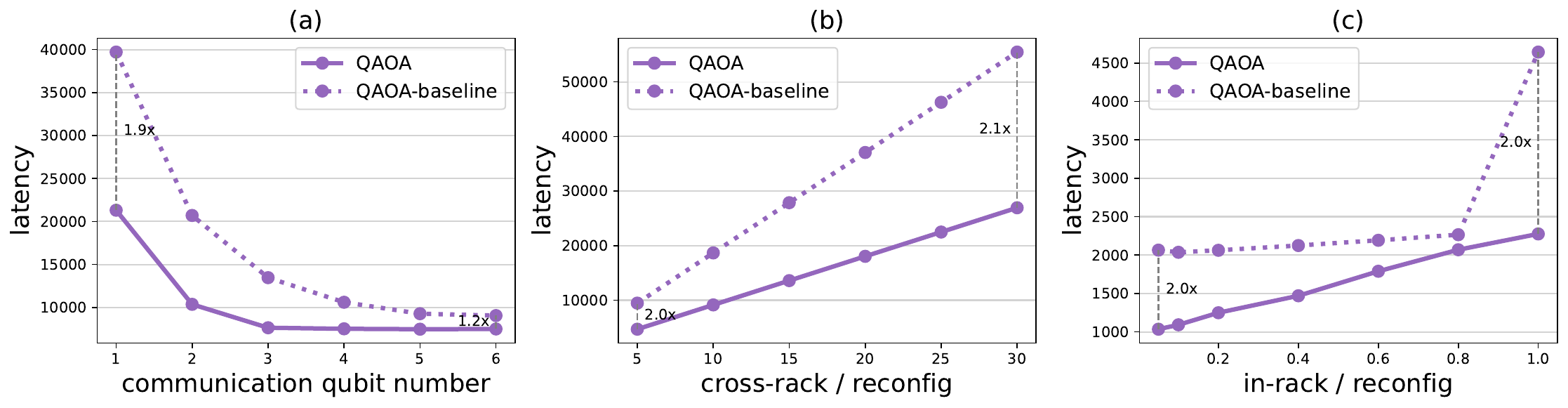}
        \caption{QAOA sensitivity varying with (a) communication qubits per QPU, (b) cross-rack EPR latency normalized by reconfiguration latency, and (c) in-rack EPR latency normalized by reconfiguration latency.}
        \label{fig:fig9_qaoa}
\vspace{-1em}
\end{figure*}

\subsection{Parameter Sensitivity}

We next evaluate how QAOA responds to compiler and hardware parameters under Program-480. Figure~\ref{fig:fig8_qaoa} shows that increasing buffer size reduces latency until the benefit saturates near 11 buffer qubits, corresponding to approximately 26.8\% of the total qubits per QPU under the evaluated configuration. This suggests that QAOA exhibits moderate buffer sensitivity under the evaluated configuration. It benefits from storing near-future EPR pairs, but its graph-dependent interactions do not require the same buffering behavior as denser workloads like QFT or Grover's from the existing list of benchmarking programs that are part of SwitchQNet. Increasing look-ahead depth produces a similar trend, with most of the benefit obtained at shallow-to-moderate depths and only small fluctuations at larger depths.

Figure~\ref{fig:fig9_qaoa} reports sensitivity to communication-qubit count and EPR generation latency. Increasing the number of communication qubits reduces latency for both the baseline and SwitchQNet, but the relative improvement narrows as communication bandwidth becomes less constrained. Increasing cross-rack EPR latency raises absolute latency for both schedulers, while the extended SwitchQNet workflow preserves an approximately 2$\times$ advantage over the baseline. In-rack EPR latency has a weaker effect, consistent with the observation that QAOA performance is dominated more by cross-rack entanglement availability than by local in-rack generation time.

We also vary EPR fidelity assumptions to estimate overhead sensitivity. As cross-rack fidelity approaches in-rack fidelity, the weighted EPR overhead increases because additional in-rack EPR pairs used for latency hiding become relatively more visible in the overhead calculation. Conversely, improving distilled in-rack fidelity reduces the overhead associated with post-split EPR generation. These results suggest that QAOA remains compatible with SwitchQNet's latency-hiding mechanisms, but its lower communication density limits the magnitude of gains compared with QFT, Grover, and RCA.

\subsection{QEC Integration}
We also evaluate QAOA using SwitchQNet's surface-code-based QEC model with Program-480 and code distance $d=5$. Under this setting, QAOA obtains a latency reduction from 11,233 to 7,462 normalized time units, corresponding to a 1.51$\times$ improvement, with 12.00\% EPR overhead and retry overhead of 1.00. The smaller gain compared with communication-heavy benchmarks is consistent with QAOA's graph-dependent interaction pattern, which provides fewer opportunities for aggressive latency hiding and collective EPR generation. The result indicates that SwitchQNet's network-aware scheduling remains beneficial for QAOA even under fault-tolerant communication assumptions.

\section{Conclusion}

This paper examines the behavior of QAOA in a DQC setting using the SwitchQNet compiler framework. Motivated by the scalability limitations of monolithic quantum hardware and the growing relevance of modular and data-center-style quantum architectures, the work addressed the absence of representative optimization workloads as a key gap in existing evaluations of distributed quantum compilers. By integrating QAOA into SwitchQNet's underlying routing and scheduling mechanisms, this study enabled a consistent analysis of how a structured, graph-based quantum combinatorial optimization algorithm interacts with quantum data center topologies. The experimental results demonstrate that QAOA exhibits distinct performance trends compared to previously studied benchmark programs, particularly with respect to look-ahead depth, communication qubit availability, and interconnect latency (cross-rack switches). The 3-regular QAOA exhibits lower absolute communication demand and modest improvement factors, complementing SwitchQNet by showing that QAOA stresses the compiler differently, exposing how graph-dependent interactions limit opportunities for collective EPR generation while still benefiting from look-ahead scheduling and entanglement-aware resource management. The sensitivity analysis further shows that moderate increases in entanglement resource consumption, including additional EPR generation and distillation, may effectively mitigate communication delays without introducing prohibitive overhead. These effects are observable even at the modest system scales explored in simulation, indicating that network-aware compilation decisions could meaningfully influence performance well before fault-tolerant regimes are reached. Collectively, the results support the central conclusion that distributed quantum algorithm performance must be evaluated through a joint consideration of algorithmic structure, network topology, and entanglement management, which will play an impactful role in the implementation of hardware architecture co-design and compilation strategies. By extending SwitchQNet with QAOA and clarifying the correspondence between its conceptual model and implementation, this work establishes a reproducible foundation for future studies of optimization algorithms in distributed quantum environments. More broadly, it positions QAOA as both a practical workload and an effective diagnostic tool for guiding the co-design of quantum algorithms, network architectures, and compilers in emerging quantum data center systems.

\section*{Acknowledgment}
The author used OpenAI ChatGPT to assist with portions of code debugging, language editing, organization, consistency checks, and formatting for selected manuscript sections. The author reviewed and verified all technical claims, experimental results, citations, figures, and final text. We thank the SwitchQNet authors for making their code open-source~\cite{zhang2025zenodo}.

\vspace{12pt}

\end{document}